# Theoretical Evaluation of Li et al.'s Approach for Improving a Binary Watermark-Based Scheme in Remote Sensing Data Communications

Mohammad Reza Khosravi [1], Mohammad Kazem Moghimi [2]
1- Department of Electrical and Electronic Engineering, Shiraz University of Technology (SUTech), Shiraz, Iran
2- Department of Electrical Engineering, Najafabad Branch, Islamic Azad University, Najafabad, Isfahan, Iran
Email: m.khosravi@sutech.ac.ir, moghimi.kazem@sel.iaun.ac.ir (Corresponding author: M. K. Moghimi)

**ABSTRACT:**
This letter is about a principal weakness of the published article by Li *et al.* in 2014. It seems that the mentioned work has a terrible conceptual mistake while presenting its theoretical approach. In fact, the work has tried to design a new attack and its effective solution for a basic watermarking algorithm by Zhu *et al.* published in 2013, however in practice, we show the Li *et al.*'s approach is not correct to obtain the aim. For disproof of the incorrect approach, we only apply a numerical example as the counterexample of the Li *et al.*'s approach.

**KEYWORDS:** Remote sensing, Digital watermarking, Counterexample, Li *et al*.'s approach for Zhu *et al*.'s watermarking algorithm.

## 1. INTRODUCTION

In 2013, Zhu *et al.* have presented their new watermarking algorithm for copyright protection of remote sensing images [1] which have specific features in terms of intensity, edge, texture and so on [3-8] (e.g. the case in Figure 1).

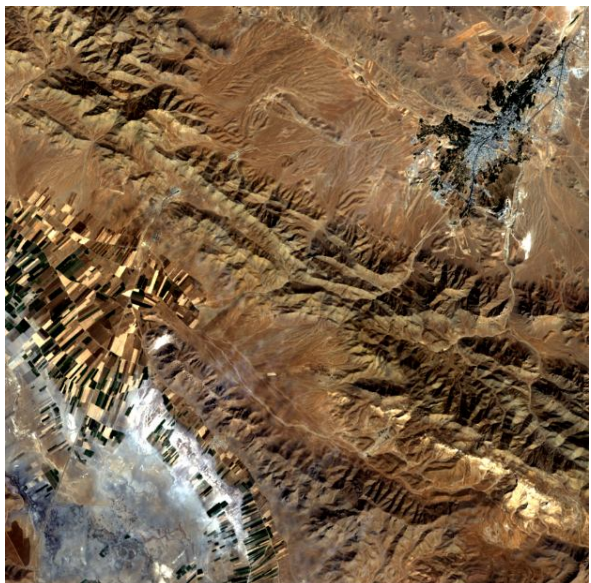

Figure 1. A sample for remote sensing images [4,8]

In 2014, Li *et al.* [2] have introduced a modification for Zhu *et al.*'s scheme. In fact, Li *et al.* firstly tried to find a security hole in Zhu *et al.*'s scheme and then proposed a certain attack for this scheme. And finally, they solved the problem regarding to this attack by introducing a solution based on a chaotic hash function. In this letter, Li *et al.*'s work was strongly disproved because we believe discovering the security hole in their work has a problem and in the other words, there is no hole in the Zhu *et al.*'s scheme. Thus, all modifications done for improving the basic algorithm are principally unnecessary.

## 2. THEORETICAL EVALUATION

According to third part in [2], in the original scheme [1,3] although all of the pixels in the host image I are summed, the final achieved check bit $b$ is just a bit. To make $b$ equal to binary watermark W which is a bit of the encrypted watermark to be embedded, at most one pixel in the image will be reserved, as stated in fifth step of watermark embedding process of the original image [2]. It means that the embedded binary watermark W can be changed by varying just one pixel of the watermarked image $I_○$ which opens a security hole. Details of the designed attack for this aim are as below [2]:

*Step1*: For the image I (or the block $I_{i,j}$ in [1], equivalently), compute $\text{sum}(I) \mod 2$ to obtain an original binary check bit $b$.

*Step2*: For the image $\tilde{I}$ (the attacker image) in which another watermark $\tilde{W}$ is embedded in it, compute $\text{sum}(\tilde{I}) \mod 2$ to obtain another binary check bit $\tilde{b}$.

*Step3*: If $b$ is equal to $\tilde{b}$, do nothing; else, reverse (do complementation) each pixel in I.

According to [2] and after performing the above steps, the watermark W embedded in original watermarked image is replaced with $\tilde{W}$ of the attacker's image. In [2] and after these steps, the authors have done some proofs and presented solutions for the attack problem, however, they have accepted steps 1 to 3 as a lemma without any proof and exactly this case creates a weakness, because accuracy of these steps is not considered. We are going to disprove it by a solved numerical example. If these steps become disproved, then all concluded items based on these approaches will be disproved.

Assume that we have a $2 \times 2$ host image with depth 2-bit as follow:

$$I = \begin{pmatrix} 0 & 1 \\ 2 & 3 \end{pmatrix} \quad (1)$$

And the attacker's image is the same of original host image, this issue is not forbidden according to [2], but we can select $\tilde{I}$ different from I.

$$\tilde{I} = I = \begin{pmatrix} 0 & 1 \\ 2 & 3 \end{pmatrix} \quad (2)$$

Also, assume original binary watermark W and the attacker's binary watermark $\tilde{W}$ are as follow:

$$W = \begin{pmatrix} 0 & 0 \\ 1 & 1 \end{pmatrix}$$
$$\tilde{W} = \begin{pmatrix} 1 & 1 \\ 1 & 1 \end{pmatrix} \quad (3)$$

According to step (3) of attack process, an enough way for attack is $b = \tilde{b}$. We now see with using two images I and $\tilde{I}$ as Eq. (1) and Eq. (2); $b = \tilde{b} = 6 \mod 2 = 0$. Therefore, we complete embedding process and then extraction process and finally will see in this example that W will not be replaced with $\tilde{W}$, against step (3). If $I_\circ$ and $\tilde{I}_\circ$ are watermarked images based on $\{I, W\}$ and $\{\tilde{I}, \tilde{W}\}$, respectively, then they are computable using embedding process of the original scheme [1], as below:

$$I_\circ = \begin{pmatrix} 0 & 1 \\ 1 & 0 \end{pmatrix}$$
$$\tilde{I}_\circ = \begin{pmatrix} 3 & 2 \\ 1 & 0 \end{pmatrix} \quad (4)$$

As a result, with using $I_\circ$, $\tilde{I}_\circ$ and check bits $b = \tilde{b} = 0$, we perform extraction process of the original scheme and compute extracted watermarks $W_{\text{extracted}}$ and $\tilde{W}_{\text{extracted}}$.

$$\left. \begin{array}{l} W_{\text{extracted}} = \begin{pmatrix} 0 & 0 \\ 1 & 1 \end{pmatrix} = W \\ \tilde{W}_{\text{extracted}} = \begin{pmatrix} 1 & 1 \\ 1 & 1 \end{pmatrix} = \tilde{W} \end{array} \right\} \quad (5)$$

$$\Rightarrow W_{\text{extracted}} \neq \tilde{W}_{\text{extracted}}$$

## 3. CONCLUSION

Thus by solving a numerical example, we showed W was not replaced with $\tilde{W}$; in the other words, we disproved the Li *et al.*'s approach for improving the Zhu *et al.*'s watermarking scheme only by a counterexample.

**REFERENCES**
[1] P. Zhu, F. Jia, J. Zhang, **"A copyright protection watermarking algorithm for remote sensing image based on binary image watermark,"** Optik-International Journal for Light and Electron Optics, vol. 124, pp. 4177–4181, 2013.
[2] M. Li, J. Zhang, W. Wen, **"Cryptanalysis and improvement of a binary watermark-based copyright protection scheme for remote sensing images,"** Optik-International Journal for Light and Electron Optics, vol. 125, 2014.
[3] M. R. Khosravi, H. Rostami, **"Improving the Binary Watermark-based Data Hiding Scheme in Remote Sensing Images,"** ICAUCAE 2016, Tehran, Iran, 2016.
[4] M. R. Khosravi, A. Keshavarz, H. Rostami, S. Mansouri, **"Statistical Image Fusion for HR Band Colorization in Landsat Sensors,"** 20th Annual Conference of Computer Society of Iran (CSICC2015), CSI Conference Publications, FUM, Mashhad, Iran, pp. 245-250, 2015.
[5] M. R. Khosravi, A. Khosravi, M. Shadloo, A. Keshavarz, **"A Novel Fake Color Scheme Based on Depth Protection for MR Passive/Optical Sensors,"** IEEE2015-The Second International Conference on Knowledge-Based Engineering and Innovations (KBEI), IUST, Tehran, Iran, pp. 362-367, 2015.
[6] M. R. Khosravi, A. Khosravi, M. Shadloo, A. Keshavarz, **"A Novel Fake Color Scheme Based on Depth Protection for MR Passive/Optical Sensors,"** Journal of Informatics and Computer Engineering, vol. 1(1), pp. 35-40, 2015.

[7] M. R. Khosravi, H. Rostami, **"A New Statistical Technique for Interpolation of Landsat Images,"** ICAUCAE 2016, Tehran, Iran, 2016.

[8] M. R. Khosravi, S. Mansouri, A. Keshavarz, H. Rostami, **"MRF-based multispectral image fusion using an adaptive approach based on edge-guided interpolation,"** Computer Science - Part. Computer Vision and Pattern Recognition, arXiv/1512.08475, December 2015.